\documentclass[fleqn,twoside]{article}
\usepackage{espcrc2}

\usepackage{graphicx}
\usepackage[figuresright]{rotating}

\newcommand{\AmS}{{\protect\the\textfont2
  A\kern-.1667em\lower.5ex\hbox{M}\kern-.125emS}}

\title{The aperiodic timing behaviour of the accretion--driven millisecond pulsar SAX J1808.4--3658}

\author{Steve van Straaten\address[API]{Astronomical Institute 
        ``Anton Pannekoek'',
        University of Amsterdam,\\  
Kruislaan 403, 1098 SJ Amsterdam, The Netherlands.}%
        ,
        Michiel van der Klis\addressmark[API]
        and
        Rudy Wijnands\address{School of Physics and Astronomy, University of St Andrews, \\
        North Haugh, St Andrews, Fife, KY16 9SS, Scotland, UK.}
 }

\begin{document}

\begin{abstract}
We studied the aperiodic X-ray timing behaviour of the accreting millisecond 
pulsar SAX J1808.4--3658. The source was recently found \cite{wijnands03} to 
be the first accreting millisecond pulsar that shows the kilohertz quasi-periodic 
oscillations (kilohertz QPOs) that are found in many other X-ray binaries with accreting neutron stars.
The high frequency of these signals reflects the short dynamical time scales in the 
region near the compact object where they originate. We find that in addition to the kilohertz QPOs
SAX J1808.4--3658 shows several low frequency timing features, based on which the source can be 
classified as a so--called atoll source.
The frequencies of the variability components of the atoll sources follow a universal scheme 
of correlations. The correlations in SAX J1808.4--3658 are similar but show a shift in upper 
kilohertz QPO frequency. This discrepancy is perhaps related to a stronger or differently configured 
magnetic field. 
\vspace{1pc}
\end{abstract}

\maketitle

\section{INTRODUCTION}

Many of the neutron star low--mass X--ray binaries can be classified as so--called 
atoll sources, based on the correlated behaviour of their timing properties at low 
frequencies ($\nu < 200$ Hz) and their X--ray spectral properties \cite{hk89}.
The atoll sources show quasi-periodic oscillations (QPOs) with frequencies ranging from a 
few hundred Hz to more than 1000 Hz (kilohertz QPOs). The low--frequency part of the power 
spectra is usually dominated by a broad band--limited noise component. In addition to 
the band--limited noise the atoll sources show several Lorentzian components below 200 Hz
(see e.g. \cite{vstr02}). In the atoll sources all components become broader as their 
characteristic frequency decreases \cite{pbk99,vstr02}. Therefore several features 
can only be classified as a QPO (FWHM $<$ centroid frequency/2) at high frequencies.
In addition to these timing features, that are observed in the persistent emission of the systems,
11 of the atoll sources show millisecond oscillations (270--619 Hz) during thermonuclear X-ray bursts 
(see \cite{strohmayer01} for a review). These burst oscillations are thought to be connected with the
spin of the neutron star and were the first evidence that atoll sources harbour millisecond pulsars.

In 1998 the first accreting millisecond pulsar was found with the Rossi X--ray Timing Explorer (RXTE)
during an outburst of SAX J1808.4--3658 \cite{wijnands98a}. During this outburst no thermonuclear X-ray bursts
were observed and the source did not show traditional kilohertz QPOs \cite{wijnands98b}. The low frequency
power spectrum was very similar to that of other atoll sources \cite{wijnands98b} and both the atoll sources 
and SAX J1808.4--3658 followed one relation when the characteristic frequency of a low--frequency Lorentzian was 
plotted versus that of the band--limited noise \cite{wijnands99}. During a new outburst of SAX J1808.4-3658
in October 2002, for the first time kilohertz QPOs and burst oscillations were observed in an accreting 
millisecond pulsar \cite{wijnands03,chakrabarty03}. Recently, four additional accreting millisecond pulsars
were seen. Those systems (and the 2002 outburst of SAX J1808.4--3658) are reviewed in detail elsewhere in
these proceedings \cite{wijnands_bepposax}. Here we present a preliminar analysis of the aperiodic--timing 
behaviour of SAX J1808.4--3658 in the frequency range 0.004--4096 Hz.
We study the correlations between the characteristic frequencies of the different timing features and compare 
the behaviour of SAX J1808.4--3658 with that of three well studied atoll sources (for a more detailed presentation
we refer to \cite{vstr_inprep}).

\section{ANALYSIS} 

In this analysis we study the 1998 and 2002 outbursts of SAX J1808.4-3658 using data from RXTE's 
proportional counter array (PCA). The 
data are divided into observations that consist of one to several satellite orbits. We exclude 
data for which the angle of the source above the Earth limb is less than 10 degrees or for which
the pointing offset is greater then 0.02 degrees.

We construct power spectra per observation using data segments of 256 s and 1/8192 s time bins 
such that the lowest available frequency is 1/256 Hz and the Nyquist frequency 4096 Hz; the 
normalization of Leahy et al. \cite{lea83} was used. We subtracted a Poisson noise level using the method 
of Klein--Wolt et al. \cite{kleinwolt03}, which is built on the analytical function from Zhang et al. \cite{zhang95}.
The resulting power spectra were then converted to squared fractional rms. 
To improve the statistics several observations, adjacent in time, and for which the power spectra 
remained the same, were added up. 
Before fitting, all frequency bins containing the 401 Hz pulse spike were removed from the power spectra.

As a fit function we use the multi--Lorentzian function; a sum of 
Lorentzian components (see e.g. \cite{bpk02,vstr02}). We only 
include those Lorentzians in the fit whose power could be measured to an accuracy of better than 33\%.
We plot the power spectra in the power times frequency representation ($\nu{\rm P}_{\nu}$), where the 
power spectral density is multiplied by its Fourier frequency. For a multi--Lorentzian fit function 
this representation helps to visualize a characteristic frequency, $\nu_{\rm max}$, namely, the frequency 
where each Lorentzian component contributes most of its variance per logarithmic frequency interval: in 
$\nu{\rm P}_{\nu}$ the Lorentzian's maximum occurs at $\nu_{\rm max}$
($\nu_{\rm max} = \sqrt{\nu_0^2 + \Delta^2}$, where $\nu_0$ is the centroid and $\Delta$ the HWHM 
of the Lorentzian \cite{bpk02}).

\section{RESULTS \& DISCUSSION}

We find that four to six Lorentzian components were needed to fit the power spectra
of SAX J1808.4-3658. The power spectra are very similar to those of other atoll sources, and all
Lorentzian components could be identified in the scheme of Belloni et al. \cite{bpk02} and 
van Straaten et al. \cite{vstr03}. 

\begin{figure}[h]
\includegraphics[width=17pc]{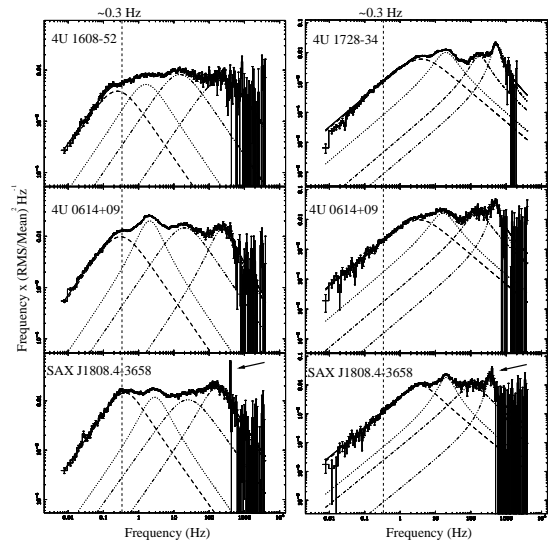}
\caption{Power spectra of SAX J1808.4--3658 and the atoll sources 4U 0614+09, 
4U 1728--34 and 4U 1608--52. Going from left to right (thus increasing in frequency), the left panel 
shows L$_u$, L$_\ell$, L$_h$, and L$_b$ when they are at the lowest characteristic 
frequencies observed for these sources. The SAX J1808.4--3658 power spectrum is from the 1998 outburst. 
The right panel shows L$_u$, L$_{hHz}$, L$_h$, and L$_b$ all at higher frequencies 
than in the left panel. The SAX J1808.4--3658 power spectrum is from the 2002 outburst. 
The arrows indicate the pulsar spike that was excluded during the fit. 
The vertical dashed line at ~0.3 Hz, approximately the 
characteristic frequency of L$_b$ in the left panel, is to guide the eye.}
\label{fig:powerspectra}
\end{figure}

An extensive description of this scheme can be found in van Straaten et al. \cite{vstr03}. To summarise;
L$_u$ is the upper kilohertz QPO. L$_\ell$ is the lower kilohertz QPO 
at high frequencies ($\nu > 200$ Hz) and a broad bump at low frequencies which does not appear 
simultaneously. Note, however, that it is unclear whether these two features are actually the same component. 
The hectohertz Lorentzian,
a broad Lorentzian with a frequency that is nearly constant and close to 150 Hz is called L$_{hHz}$.
We call the band--limited noise component that ``transforms'' into a QPO at higher frequencies (above 15 Hz) as 
well as the QPO it becomes L$_{b}$ and the ``new'' broad band--limited noise appearing at lower frequency 
L$_{b2}$, but please note that there are uncertainties in the interpretation underlying this nomenclature.
Finally there is another low frequency Lorentzian above L$_{b}$ called L$_h$.
In Figure \ref{fig:powerspectra} we show two power spectra of SAX J1808.4--3658 compared to those 
of other atoll sources studied in \cite{vstr02,vstr03}. 

\begin{figure}[h]
\includegraphics[width=14pc]{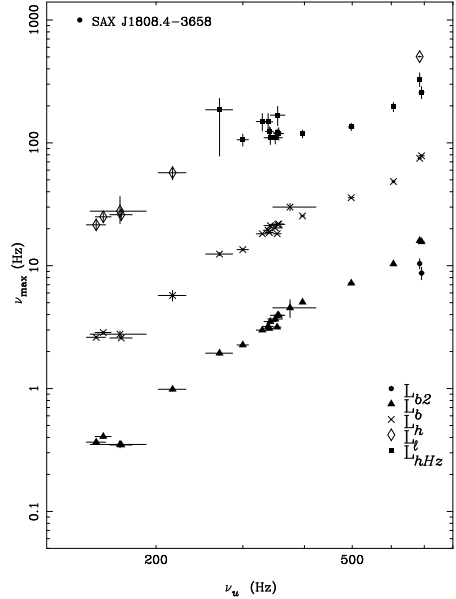}
\caption{Correlations for SAX J1808.4--3658 between the characteristic frequencies 
of the several power spectral components (names according to the scheme of \cite{bpk02,vstr03} are
indicated in the plot) and the characteristic frequency of the Lorentzian identified as the upper 
kilohertz QPO.}
\label{fig:freq_freq_sax}
\end{figure}

The frequencies of the variability components of the atoll sources 
follow a universal scheme of correlations, suggesting a very similar 
accretion flow configuration \cite{vstr02,vstr03}. If we plot the 
characteristic frequencies of the several power spectral components
versus that of the Lorentzian identified as the upper kilohertz QPO,
SAX J1808.4--3658 shows similar correlations (see Figure 
\ref{fig:freq_freq_sax}). 

\begin{figure}[h]
\includegraphics[width=15pc]{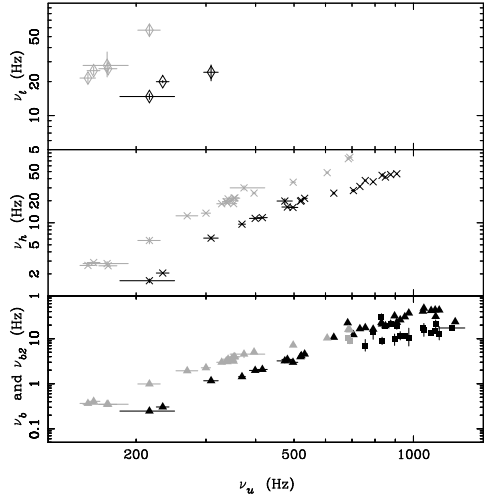}
\caption{Characteristic frequencies of the low frequency (see above) 
L$_\ell$ (top panel), L$_h$ (middle panel), L$_{b}$ and L$_{b2}$ 
(lower panel) plotted versus the 
characteristic frequency of the Lorentzian identified as the upper 
kilohertz QPO. The grey points are for SAX J1808.4--3658 and the black point are for the atoll sources 
4U 0614+09, 4U 1728--34, and 4U 1608--52. The SAX J1808.4--3658 
relations are shifted to lower $\nu_u$.}
\label{fig:freq_freq_all}
\end{figure}

We can use figure \ref{fig:freq_freq_sax} to identify
the upper kilohertz QPO (L$_u$) at the lowest characteristic
frequencies ($\approx$ 150 Hz) found in any source to date. 
These characteristic frequencies were reached during the 1998 outburst where  
the characteristic frequencies of all the power spectral components were generally 
lower than during the 2002 outburst.
Previous work \cite{wijnands98b} already did observe 
these features, but did not recognize them as upper kilohertz QPOs due to 
their low frequencies and broad character. Now that kilohertz QPOs have been observed at high 
frequencies in SAX J1808.4--3658, the correlations in figure \ref{fig:freq_freq_sax}  
clearly show that the upper kiloherz QPO is present in a broad frequency range.

If we compare the relations of SAX J1808.4--3658  with those of the atoll sources
4U 0614+09, 4U 1728--34 \cite{vstr02}, and 4U 1608--52 \cite{vstr03} we find similar 
correlations, but we see that the upper kilohertz QPO frequency is too low by
a factor of about 1.5 compared to the atoll sources (see Figure \ref{fig:freq_freq_all}).

A possible explanation for the shift in upper kilohertz QPO frequency
could be that the magnetic field strenght $B$ is higher in SAX J1808.4--3658:
In most models the upper kilohertz QPO frequency reflects the Keplerian frequency 
at the inner edge of the accretion disk (e.g. \cite{miller98}).
Because of the weaker $B$ field in the atoll sources the inner radius of 
the disk can get closer to the neutron star surface. If the low frequency features 
are formed further out in the disk their radii and therefore frequencies  
are independent of $B$ field strenght. So while the low frequency features have similar 
characteristic frequencies in the atoll sources as in SAX J1808.4--3658, the upper 
kilohertz QPO frequency is lower.

\section{CONCLUSIONS}

\begin{itemize} 
 
\item        SAX J1808.4-3658 can be classified as an atoll source based 
	     on its low frequency timing.

\item        The upper kilohertz QPO of Wijnands et al. \cite{wijnands03} is present in a wide (150--700 Hz) 
	     frequency range. 
	     At the lowest characteristic frequencies, observed during the 1998 outburst, 
	     the features were previously not recognised as upper kilohertz QPO, but could now be identified 
	     with the help of frequency relations.
	     
\item        The first accreting milisecond pulsar plotted on the frequency relations diagram 
	     deviates from the other sources. The deviation is most easily described as a shift in upper 
	     kilohertz QPO frequency. This shift might be due to a higher $B$ field strength
             in SAX J1808.4-3658.

\end{itemize}

\end{document}